\input harvmac
\input epsf

\newcount\figno
\figno=0
\def\fig#1#2#3{
\par\begingroup\parindent=0pt\leftskip=1cm\rightskip=1cm\parindent=0pt
\baselineskip=11pt
\global\advance\figno by 1
\midinsert
\epsfxsize=#3
\centerline{\epsfbox{#2}}
\vskip 12pt
\centerline{Fig. \the\figno. #1}\par
\endinsert\endgroup\par}

\font\mbm=msbm10
\def\bb#1{\hbox{\mbm #1}}

\lref\ggs{A. Sagnotti, Phys. Lett. B294 (1992) 196; \hfill\break
S. Ferrara, R. Minasian and A. Sagnotti, Nucl. Phys. B474 (1996)
323;\hfill\break
S. Ferrara. F. Riccioni and A. Sagnotti, Nucl. Phys. B519 (1998)
115; \hfill\break
F. Riccioni and A. Sagnotti, Phys. Lett. B436 (1998) 298.}

\lref\bgkfut{R. Blumenhagen, L. G\"orlich and B. K\"ors, in preparation.}

\lref\toroidal{M. Bianchi, G. Pradisi and A. Sagnotti,
Nucl. Phys. B376 (1992) 365.}

\lref\bs{M. Bianchi and A. Sagnotti, Phys. Lett. B247 (1990) 517;
Nucl. Phys. B361 (1991) 519.}

\lref\comm{Z. Kakushadze, G. Shiu and S.-H.H. Tye, 
Phys. Rev. D58 (1998) 086001;\hfill\break
C. Angelantonj, {\it Comments on open string orbifolds with
a non-vanishing $B_{ab}$}, hep-th/9908064.}

\lref\bgk{R. Blumenhagen, G\"orlich and B. K\"ors, {\it Supersymmetric 
orientifolds in 6D with D-branes at angles}, hep-th/9908130.}

\lref\cargese{A. Sagnotti, in Carg{\'e}se 87, ``Non-Perturbative Quantum Field
Theory'', eds. G. Mack et al. (Pergamon Press, Oxford, 1988), p. 521.}

\lref\bimopra{M. Bianchi, J.F. Morales and G. Pradisi, {\it
Discrete torsion in non-geometric orbifolds and their open-string
descendants}, hep-th/9910228.}

\lref\bs{M. Bianchi and A. Sagnotti, Phys. Lett. B247 (1990) 517;\hfill\break
M. Bianchi and A. Sagnotti, Nucl. Phys. B361 (1991) 519.}

\lref\bg{R. Blumenhagen and L. G{\"o}rlich, Nucl. Phys. B551 (1999) 601.}

\lref\gepner{C. Angelantonj, M. Bianchi, G. Pradisi, A. Sagnotti and
Ya.S. Stanev, Phys. Lett. B387 (1996) 743.}

\lref\bw{R. Blumenhagen and A. Wi{\ss}kirken, Phys.Lett. B438 (1998) 52.}


\Title{
\vbox{
\rightline{CPHT-S744.1099}
\rightline{HUB-EP-99/60}
\rightline{LPTENS 99/38}
\rightline{\tt hep-th/9911190} 
\vskip -.5in}}
{Discrete Deformations in Type I Vacua}
\vskip -.25in
\centerline{Carlo Angelantonj${}^1$ and Ralph Blumenhagen${}^2$} 
\medskip
\centerline{\it ${}^1$ Centre de Physique
Th{\'e}orique\footnote{${}^\dagger$}{Unit\'e mixte du CNRS et de l'EP,
UMR-7644}, 
{\'E}cole Polytechnique, F-91128 Palaiseau}
\centerline{\it and}
\centerline{\it Laboratoire de Physique 
Th{\'e}orique de l'{\'E}cole Normale 
Sup{\'e}rieure\footnote{${}^\ddagger$}{Unit{\'e} mixte du 
CNRS et de l'ENS, UMR 8519},}
\centerline{\it 24 rue Lhomond, F-75231 Paris Cedex 05}
\centerline{\it ${}^2$ Humboldt-Universit\"at zu Berlin, Institut f\"ur 
Physik,}
\centerline{\it Invalidenstrasse 110, 10115 Berlin, Germany}

\vskip 0.3in

\centerline{{\bf Abstract}}

\noindent
We study supersymmetric orientifolds where the world-sheet parity 
transformation is combined with a conjugation of some
compact complex coordinates. We investigate their T-duality relation 
to standard orientifolds and discuss the origin of continuous and 
discrete moduli. 
In contrast to standard orientifolds, the antisymmetric tensor describes
a continuous deformation, while the off-diagonal part of the
metric is frozen to quantized values and  is responsible for the rank
reduction of the gauge group. We also give a geometrical
interpretation of some recently constructed six-dimensional 
permutational orientifolds.

\Date{11/99} 


\newsec{Introduction}

Since the early work of Bianchi, Pradisi and Sagnotti \toroidal,
it is known that Type I
string vacua involve both continuous and discrete moduli.  
Thus, in the case of ordinary orientifolds based on the world-sheet 
parity $\Omega$ \cargese, discrete internal backgrounds for the NS-NS
two-form, $B_{ab}$, are still allowed, although the corresponding (continuous)
deformations are projected out.
A background $B$-field generically leads to a rank reduction
for the gauge group, while, in the case of orbifold
compactifications, extra multiplicities appear in some
open string sectors \comm.  

In a recent paper \bgk\ a new class of six-dimensional
orientifolds of the type IIB string has been studied. They involve
the combined action of the world-sheet parity $\Omega$ with a conjugation
\bg\ ${\cal R}$
of the internal complex coordinates and a geometric $\bb{Z}_N$ action.
The new feature arising in these orientifolds is that the Klein
bottle amplitude receives contributions from all $\bb{Z}_N$ twisted
sectors. After an $S$ transformation to
the transverse tree-channel, these lead to new tadpoles whose cancellation
requires (different types of) D7-branes 
intersecting at non-trivial (quantized) angles. Then,
strings stretched between different D7-branes give rise to new
twisted open sectors crucially involved in tadpole cancellations. 
The resulting models have ${\cal N} =(1,0)$ space-time supersymmetry in $D=6$ 
and are characterized by Chan-Paton groups of reduced rank. 
While for the standard $\Omega$ projection
this phenomenon may be related to the presence of a
quantized background for the NS-NS antisymmetric tensor
(both in toroidal \toroidal\ and in orbifold \comm\ 
compactifications), a similar understanding is missing for the
$\Omega{\cal R}$ orientifolds of \bgk . Finding an appropriate
description of the observed rank reduction is
the main motivation of the present letter.

As a simple example, in Section 2 we study 
$\Omega{\cal R}$ orientifolds of type IIA 
on a two-torus and using T-duality we gain an understanding of the 
mechanism responsible for the rank reduction.
In Section 3 we study tadpole cancellation for this model,
derive the quantization rules for geometric moduli and
show that they are consistent with the results obtained in \bgk.
Finally, in Section 4 we study the six dimensional $\bb{Z}_2$ 
$\Omega{\cal R}$ orientifold and using T-duality give an
alternative geometric interpretation of the extra multiplicities 
present in the Neuman-Dirichlet open-string sector.

\newsec{T-duality and $\Omega{\cal R}$ orientifolds}

Let us consider the type IIB string on a torus $T'^2$ with complex 
coordinate $z=x_2+{\rm i} x_1$. Without loss of generality, one can choose the
basis 
\eqn\basis{
e'_1= {\rm i} R\ , \quad\quad\quad e'_2 = - b + {\rm i} a\ ,
}
in terms of which the complex and K\"ahler structures are
\eqn\compl{ 
U' = U'_1 + {\rm i} U'_2 = {e'_2 \over e'_1} = {a \over R} + {\rm i} 
{b \over R} \ , \quad\quad
            T' = T'_1 + {\rm i} T'_2 = B' + {\rm i} V' = B' + {\rm i} b R \ ,
}
where we have also turned on an internal background $B$-field. 
It is well known that modding out by the world-sheet parity transformation 
$\Omega$, the three geometric moduli $U'_1, U'_2$, $T'_2$ survive the 
projection whereas, up to the identification $B'\equiv B'+\bb{Z}$, 
the antisymmetric tensor is frozen to the two possible
values $B'=0$ and $B'={1 \over 2}$ \toroidal . 

A T-duality along the $x_1$ direction leads to the orientifold
\eqn\tdual{ 
{{\rm type\ IIA}\ {\rm on}\ T^2\over \Omega{\cal R} }\ ,
}
where now the $\Omega$ projection is combined
with the operation ${\cal R}:\ z\to\bar z$.
Moreover, the complex and K\"ahler structures are interchanged, so that 
for the dual torus $T^2$ 
\eqn\complb{ 
U = U_1 + {\rm i} U_2 = {e_2 \over e_1} = { - b + {\rm i} B' / R 
\over {\rm i} / R}\ ,\quad\quad
            T = T_1 + {\rm i} T_2 = B + {\rm i} V = {a \over R} + 
{\rm i} {b \over R}\ .
}
Therefore, in the orientifold \tdual\ the $B$-field is 
a continuous parameter while the geometric modulus $U_1$ is frozen.
The quantization condition $2U_1 \in \bb{Z}$ expresses the fact
that for all other values of $U_1$, the action of ${\cal R}$ on $T^2$
is not crystallographic. Moreover, under T-duality 
the periodicity $B' \equiv B'+\bb{Z}$ is mapped to the invariance of the 
complex structure under integer 
${\rm SL}(2;\bb{Z})$ shifts. 

In conclusion, from 
T-duality we learn that for $2 U_1 \in 2 \bb{Z} + 1$ the rank
of the gauge group is halved. In the next 
Section,  for the 
orientifold \tdual\ we discuss tadpole cancellation in detail
and confirm the above expectations suggested by 
T-duality.   

\newsec{Tadpole cancellation for $\Omega{\cal R}$ orientifolds}

Let us therefore consider the compactification
on a generic two-torus $T^2$, whose complex ($U= U_1 + {\rm i} U_2$) and 
K\"ahler ($T= T_1 + {\rm i} T_2$) structures
are related to the metric and antisymmetric tensor by
\eqn\met{
g = {\alpha ' T_2 \over U_2} \left( \matrix{1 & U_1 \cr U_1 & |U|^2\cr} 
\right) \ , \qquad B = \alpha ' T_1 \left( \matrix{0 & -1 \cr 1 & 0\cr}
\right) \ .
}
The left and right momenta corresponding to the
complex coordinates ($z\, ,\, \bar z$) on the $T^2$ are
\eqn\plr{
\eqalign{
p_{\rm L} =& {1\over \sqrt{\alpha ' U_2 T_2}} \left[ U m_1 - m_2 - \bar T (n^1
+ U n^2) \right] \ ,
\cr
p_{\rm R} =& {1\over \sqrt{\alpha ' U_2 T_2}} \left[ U m_1 - m_2 - T (n^1
+ U n^2) \right] \ ,
\cr}
}
where $m_a$ and $n^b$ denote the KK momenta and winding numbers.
For generic values of the $U$ and $T$ moduli, the type IIB string is no more
left-right symmetric. However, one can impose the invariance under $\Omega$
and  find constraints on the allowed values of
the moduli. In particular, in the 
``conventional'' type I string, obtained projecting the parent IIB 
with respect to the simple world-sheet parity \cargese, the real slice of the
K\"ahler structure (the $B$-field) can only take discrete values ($2 T_1 \in
\bb{Z}$) \toroidal. 
This is consistent with the fact that the excitations of the
NS-NS two-form are no longer part of the physical spectrum 
of the projected theory. 
Moreover, the presence of a quantized flux for the $B$-field 
implies the reduction of the total size of 
the Chan-Paton gauge group by a factor $2^{r/2}$, with $r= {\rm rank} 
(B_{ab})$, and a continuous interpolation (via Wilson lines)
between orthogonal and symplectic gauge groups \toroidal.

Actually, one can consider more general projections and dress the world-sheet 
parity $\Omega$ with other type II symmetries as, for instance, 
coordinate reflections (T-dualities) in the compact directions. 
For the 
two-dimensional torus there exist two additional possibilities,
namely the inversion of both coordinates, 
${\cal I}:\ z \to -z$, or of a single coordinate 
\eqn\zzbar{
{\cal R}:\ z\to \bar z \ .}

It is not difficult to see that $\Omega{\cal I}$ 
does not present any substantial difference compared  to the standard 
$\Omega$ projection. 
In both cases, only the imaginary slice of the K\"ahler structure 
survives, while the resulting type I models correspond to
compactifications on T-dual tori (with D7 and D9-branes, respectively).  

However, new interesting features arise when the world-sheet parity 
is dressed with
the conjugation \zzbar . A simple analysis of the massless spectrum reveals
that the internal components of the NS-NS antisymmetric tensor
\eqn\inta{
\left(
\psi_{-{1\over 2}}^{1} \tilde\psi_{-{1\over 2}}^{2} - \psi_{-{1\over 2}}^{2} 
\tilde\psi_{-{1\over 2}}^{1} \right) |0\tilde 0\rangle
}
now survive the $\Omega{\cal R}$ projection, while the mixed components
of the internal metric
\eqn\intb{
\left(\psi_{-{1\over 2}}^{1} \tilde\psi_{-{1\over 2}}^{2} 
+\psi_{-{1\over 2}}^{2} 
\tilde\psi_{-{1\over 2}}^{1} \right) |0\tilde 0\rangle
}
do not. Therefore, in this case one expects that the antisymmetric 
tensor is a continuous modulus of the projected theory, 
while some conditions
have  to be satisfied  by the mixed components of the metric. Indeed, 
it is not hard to see that requiring invariance of the parent theory under 
\eqn\inva{
\Omega {\cal R} : \quad p_{\rm L} = \bar p_{\rm R} '
}
results in a quantization condition for the real slice of the complex
structure
\eqn\quant{
2 U_1 \in \bb{Z}\ ,
}
in analogy with the standard case. 

Once the $\Omega{\cal R}$ symmetry of the parent closed string is restored,
one can proceed as in \cargese\ to construct the open descendants.
Starting as usual from the Klein bottle amplitude 
\eqn\kba{
{\cal K} = {\textstyle{1\over 2}} \, {\rm Tr} \left( \Omega {\cal R}\,
{\cal P}_{\rm GSO}\, {\rm e}^{-2\pi t(L_0+\bar L_0)} \right)
}
one realizes that it receives contributions from the lattice states satisfying
$p_{\rm L} = \bar p_{\rm R}$, which are fixed under
$\Omega {\cal R}$. The resulting amplitude in the direct
loop-channel is
\eqn\kbb{
{\cal K} = {\textstyle{1\over 2}} (V_8 - S_8 ) (2 {\rm i} t )
\sum_{m_2 , n^1} \, 
{{\rm e}^{-{\pi t\over  U_2 T_2} |m_2 + \bar T n^1 |^2} \over \eta^2
(2{\rm i} t)} \ ,
}
and after an $S$ transformation to the transverse channel leads to the
amplitude  
\eqn\kbc{
\tilde{\cal K} 
= {2^5 \over 2} \, U_2 \, (V_8 - S_8) ({\rm i} \ell )
\sum_{m_1 , n^2} { {\rm e}^{-2\pi\ell {U_2 \over T_2} |m_1 - \bar T n^2|^2}
\over \eta^2 ({\rm i} \ell )}\ .
}
In \kbc\ we have written the contribution of the world-sheet fermions
\eqn\theta{
(1-1) \, {\vartheta_2^4 \over 2\eta^4} = V_8 - S_8
}
in terms of level-one SO(8) characters, and we have omitted the contributions
of the transverse bosonic coordinates and the integration measure.

The annulus amplitude can be computed in two ways. The first, as in
\bgk, is to place D8-branes at $x_1=0$ extending in the $x_2$ direction 
and compute the direct-channel amplitude taking care exactly
of the KK and winding states in the $x_2$ and $x_1$ directions. 
The other, followed here, is to start from the transverse
channel, that involves only the sectors of the parent theory that are 
combined with their conjugates. For the $\Omega{\cal R}$ projection,
these states are selected imposing $p_{\rm L} = -\bar p_{\rm R}$. Given the
quantization condition \quant\ on the $U_1$ field, there are further 
restrictions on these allowed states:
\eqn\quanta{
2 U_1 m_1 \,,\ 2 U_1 n^2 \in 2 \bb{Z} \ .
}
As usual, these constraints are imposed inserting in the transverse 
tree-channel annulus amplitude suitable projectors
\eqn\anna{
\tilde{\cal A} = {2^{-5 + 2a -2}\over 2} \, U_2 \, 
N^2\,(V_8 - S_8) ({\rm i}\ell )
\sum_{\epsilon_1 , \epsilon_2 =0,1} \sum_{m_1 , n^2} { {\rm e}^{-{ \pi\ell 
U_2 \over 2T_2} |m_1 - \bar T n^2 |^2} \, {\rm e}^{2{\rm i}\pi U_1 
(m_1 \epsilon_1 - n^2 \epsilon_2)} \over \eta^2 ({\rm i}\ell)} \ ,
}
where $a$ depends on the value of $U_1$, and is zero for  $U_1\in\bb{Z}$, 
and one for $U_1\in\bb{Z}+{1\over 2}$.
The coefficient $2^{2a-2}$ ensures a proper normalization of the
vector in the direct-channel amplitude
\eqn\annb{
{\cal A} = {2^{2a-2} \over 2} \, N^2 \, (V_8 - S_8) ({\rm i} t/2) 
\sum_{\epsilon_1 , \epsilon_2=0,1} \sum_{m_2 , n^1} 
{{\rm e}^{-{\pi t\over  U_2 T_2} 
|m_2 + U_1 \epsilon_2 + \bar T (n^1 + U_1 \epsilon_1)|^2} \over \eta^2 
({\rm i}t/2)} \ ,
}
that, for $U_1 \in \bb{Z} +{1\over 2}$, receives contribution only
from $\epsilon_1 = \epsilon_2 =0$.
The M\"obius amplitude
\eqn\moeba{
\eqalign{
\tilde{\cal M} 
=& - {2\times 2^{a-1} \over 2} \, U_2 \, N \, (\hat V _8 - 
\hat S _8 ) ({\rm i} \ell +{\textstyle{1\over 2}}) \times
\cr
& \times 
\sum_{\epsilon_1 , \epsilon_2=0,1} 
\sum_{m_1 , n^2} { {\rm e}^{-2 \pi\ell {U_2 \over T_2} |m_1 - 
\bar T n^2 |^2}\, {\rm e}^{2{\rm i}\pi U_1 (m_1 \epsilon_1 - n^2 \epsilon_2 )}
\, \gamma_{\epsilon_1 , \epsilon_2} \over \hat\eta^2 ({\rm i}\ell +
{\textstyle{1\over 2}})} 
\cr}}
involves signs $\gamma_{\epsilon_1 , \epsilon_2}$ as in \toroidal, 
that enforce a proper normalization of the $U_1$ projector.
Extracting the contributions of $\tilde{\cal K}$, 
$\tilde{\cal A}$ and $\tilde{\cal M}$
to the massless tadpoles, one finds the consistency condition 
\eqn\gamm{
\sum_{\epsilon_1 , \epsilon_2 = 0,1} \gamma_{\epsilon_1 , \epsilon_2}
=2 \ ,
}
and, as a result, the total size of the Chan-Paton gauge group is 
reduced by a factor $2^a$
\eqn\tad{ 
N = 2^{5-a} \ .
}
Thus, the gauge group is ${\rm SO}(2^{5-a})$ or ${\rm USp} (2^{5-a})$ 
depending on the sign of $\gamma_{0,0}$ in
\eqn\moebib{
{\cal M} =- {2^{a-1}\over 2}\, N\, (\hat V_8 - \hat S_8 ) 
\left((1 + {\rm i} t)/2\right) 
\sum_{\epsilon_1 , \epsilon_2=0,1} \sum_{m_2 , n^1} 
{{\rm e}^{-{\pi t\over  U_2 T_2} 
|m_2 + U_1 \epsilon_2 + \bar T (n^1 + U_1 \epsilon_1)|^2} \,
\gamma_{\epsilon_1,\epsilon_2} \over \hat\eta^2 
\left(( 1 + {\rm i} t)/2\right)} \ .
}
For $U_1\in\bb{Z}$
its rank is not reduced, while for
$U_1\in\bb{Z}+{1 \over 2}$ it is halved. Also in this case
Wilson lines connect continuously orthogonal and symplectic groups 
passing through unitary ones.

We can now turn to analyze more complicated cases.
In \bgk, six-dimensional $\Omega{\cal R}$ orientifolds 
with ${\cal N} =(1,0)$ space-time supersymmetry were studied. 
In particular, the orientifolds 
\eqn\sixdim{  {{\rm type\ IIB}\ {\rm on}\ T^2\times T^2
             \over \{\Omega{\cal R},\bb{Z}_N\}} }
with $N=3,4,6$ were considered. In order to define
perturbatively consistent orientifolds, the lattices had to allow a
crystallographic action of the $\bb{Z}_N$ orbifold group and
had also to be oriented in specific ways relative to
the mirror plane of  ${\cal R}$.
The massless spectra of the orientifolds \sixdim\ had the 
generic feature of rank reductions of the gauge groups by powers of two. 
For instance, for the $\bb{Z}_3$ model a natural choice was 
the lattice ${\bf A}$  depicted in Fig. 1, both in the (67) and (89) planes.
\fig{The $T^2$ lattices ${\bf A}$ for the $\bb{Z}_3$ 
orbifold.}{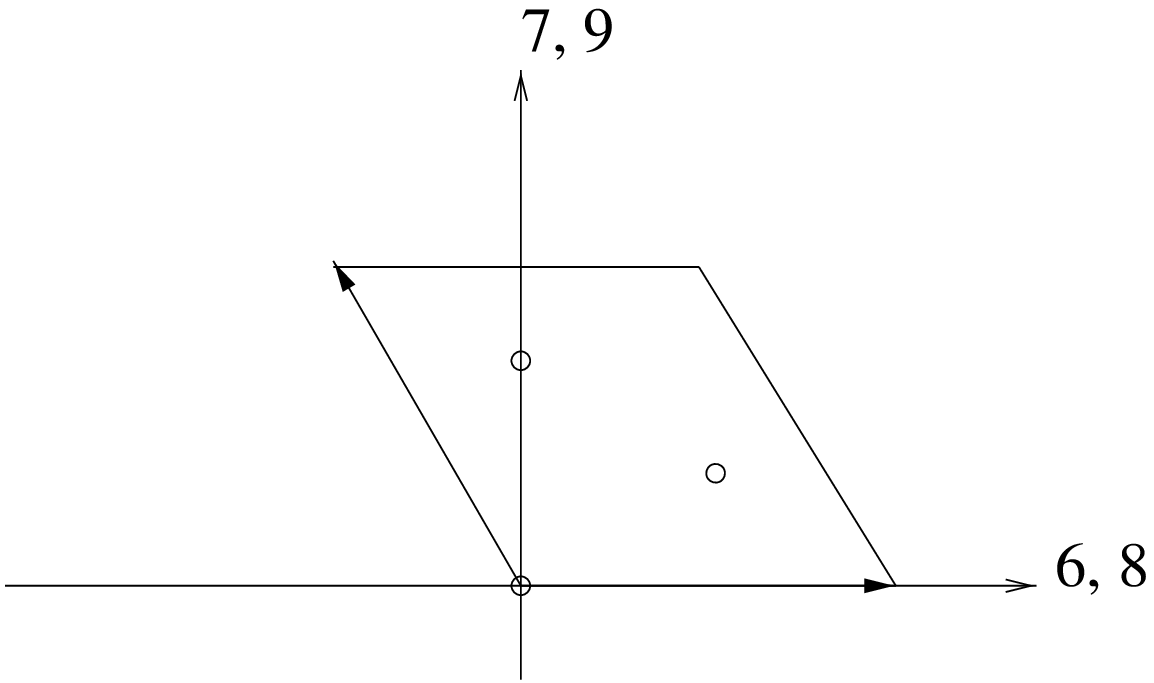}{7truecm}
Now, after applying an
${\rm SL}(2;\bb{Z})$ transformation on each $T^2$ leading to 
a different unit cell with base vectors
\eqn\unit{ 
e_1 = {\rm i} R\ , \quad\quad\quad e_2 = - {\sqrt{3}\over 2} +
{\rm i} {R\over 2}\ ,
}
one can see that the background involves a quantized value $U_1 \in \bb{Z}+{1
\over 2}$ of the real component of the complex structure. One thus expects
a rank reduction of the gauge group by a factor of two for each $T^2$,
indeed consistent with the SO(8) gauge group derived in \bgk\
for this model. For the tadpole cancellation condition and the 
computation of the massless spectrum, it is important to
know how the $\bb{Z}_3$ fixed points transform under the reflection
${\cal R}$. For the lattice in Fig. 1 one of the nine fixed points is 
invariant while the remaining eight are interchanged. 

\fig{The $T^2$ lattices ${\bf B}$ for the $\bb{Z}_3$ 
orbifold.}{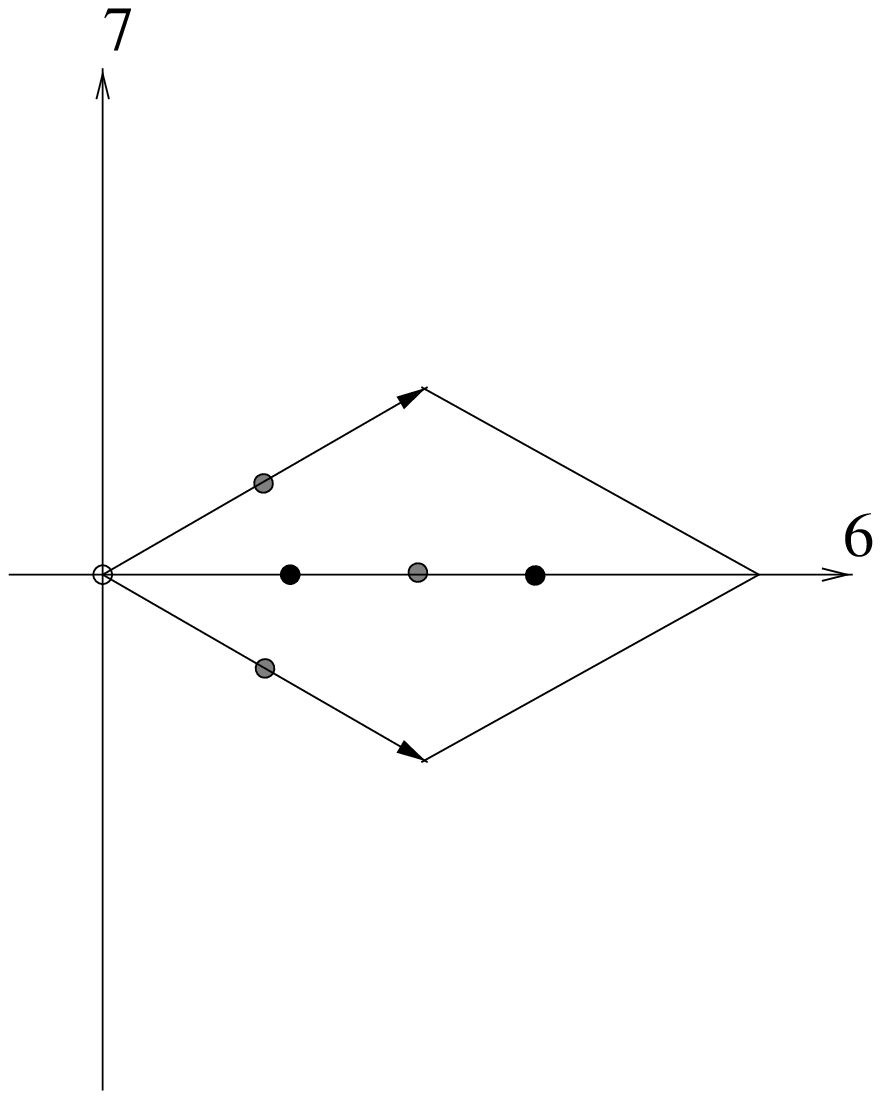}{5truecm}
Rotating the lattice by an angle $\phi=\pi/6$
results in the unit cell ${\bf B}$ of Fig. 2. By an 
${\rm SL}(2;\bb{Z})$ transformation the unit cell can be brought
to the form
\eqn\unit{ 
f_1 = {\rm i} R\ ,\quad\quad\quad f_2 = - {1\over 2\sqrt{3}}
+ {\rm i} {R\over 2}\ .
}
Note, that for the lattice ${\bf B}$ all three $\bb{Z}_3$ fixed points 
of a single $T^2$ are invariant under
${\cal R}$. Thus, besides the $T^4$ torus ${\bf AA}$ there exist two more
inequivalent choices, namely ${\bf AB}$ and ${\bf BB}$. 
Going through the computation
eventually leads to the anomaly-free massless spectra shown in Table
1. Particularly interesting is the {\bf BB} model since it corresponds 
to a dual heterotic string with a frozen dilaton.
Notice, that all these spectra ({\bf AA}, {\bf AB}, {\bf BB}) 
have already appeared  in the study of open descendants
of six-dimensional Gepner models \gepner. Moreover,  the {\bf AB} case was
recently obtained in a study of
asymmetric orientifolds \bimopra. In the Gepner construction, 
the various choices of elementary cells correspond to different modular
invariant combinations of characters for the parent closed theory. A
better understanding of this correspondence could shed some light on the
D-brane interpretation of orientifolds of Gepner models \refs{\bs,\gepner,\bw}.
\vskip 0.5cm
\centerline{\vbox{
\hbox{\vbox{\offinterlineskip
\def\tablespace{height2pt&\omit&&\omit&&
 \omit&\cr}
\def\tablerule{\tablespace\noalign{\hrule}\tablespace}

\hrule\halign{&\vrule#&\strut\hskip0.2cm\hfil#\hfill\hskip0.2cm\cr
\tablespace
& $T^4$ && closed && open       &\cr
\tablerule
& ${\bf AA}$ && $13 H\ +\ 8 T$  && $SO(8)\ +\ 2 H\ {\rm in}\ {\bf 28}$ &\cr
\tablespace
& ${\bf AB}$ && $15 H\ +\ 6 T$  && $SO(8)\ +\ 4 H\ {\rm in}\ {\bf 28}$ &\cr
\tablespace
& ${\bf BB}$ && $21 H$  && $SO(8)\ +\ 10 H\ {\rm in}\ {\bf 28}$ &\cr
\tablespace}\hrule}}}}
\centerline{
\hbox{{\bf Table 1:}{\it ~~ String spectra of $T^4/\bb{Z}_3$ }}}
\vskip 0.5cm
\noindent
For the $\bb{Z}_4$ orientifold, one had to make the choices in Figs. 3 and
4 for  the lattices in the (89) and (67) directions, that
correspond to the unit cells
\eqn\unit{
\eqalign{e_1 &= {\rm i} R_1 \ ,
\cr
f_1 &= {\rm i} R_2 \ ,
\cr}
\quad\quad\quad 
\eqalign{e_2 &= - R_1\ , \cr 
f_2 &= - {R_2 \over 2} + {\rm i} {R_2\over 2}\ . \cr }
}
\fig{The $\bb{Z}_4$ lattice ${\bf A}$ in the plane 
(89).}{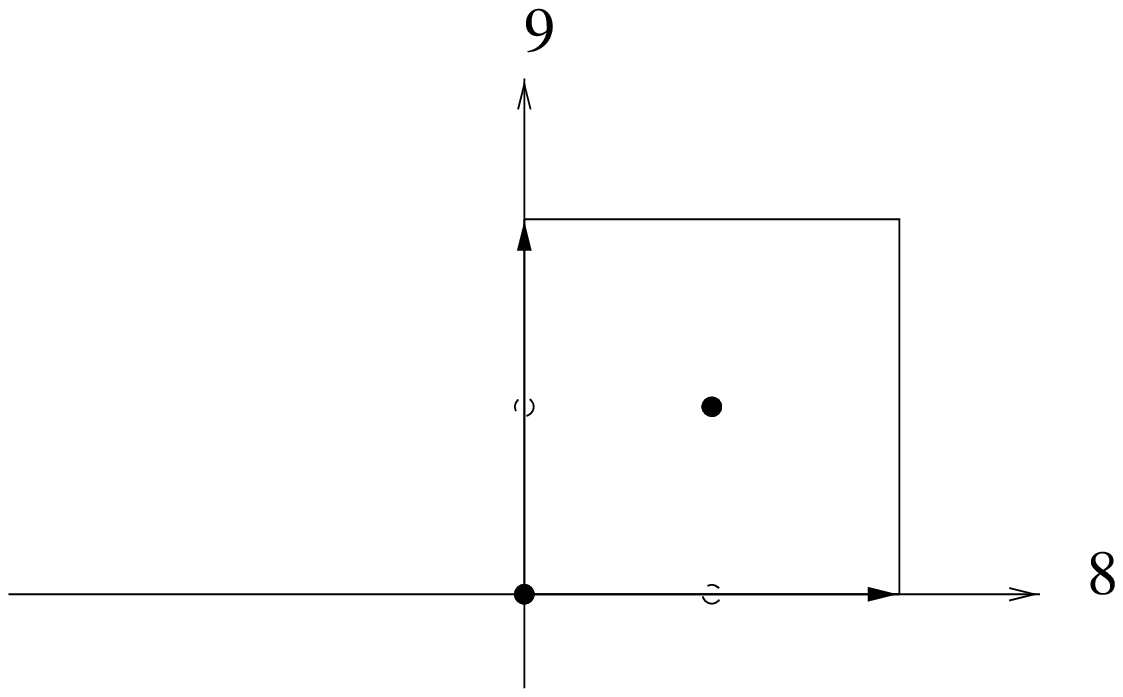}{7truecm}
\fig{The $\bb{Z}_4$ lattice ${\bf B}$ in the plane 
(67).}{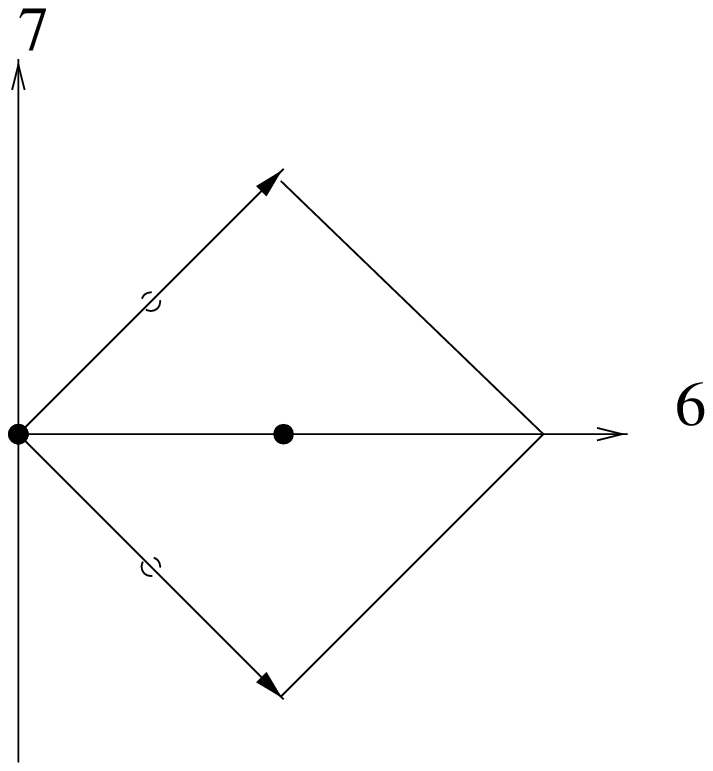}{4truecm}
\noindent
In analogy with the previous case, one would
thus expect that the rank of the gauge group is  halved for this model.
Since 
$\bb{Z}_4$ contains a $\bb{Z}_2$ subgroup, tadpole cancellation requires
the introduction of two unrelated types of D7-branes and 
the unreduced rank of the
gauge group is 32.  Consistently, the $\bb{Z}_4$ orientifold was found to 
have the gauge group ${\rm U}(8)\times {\rm U}(8)$ of rank 16.

Finally, for the $\bb{Z}_6$ orientifold the only consistent choice
of the lattices was the $\bb{Z}_3$ {\bf AB} configuration leading 
to a reduction of the gauge group by a factor of four. 
This is nicely confirmed by the resulting gauge group 
${\rm U}(4)\times {\rm U}(4)$ found in \bgk.
 
Four dimensional orientifolds with ${\cal N}=1$ supersymmetry
discussed in the forthcoming paper \bgkfut\ display gauge groups
that precisely reflect the quantization rules for  $U_1$.

\newsec{The $\bb{Z}_2$ $\Omega{\cal R}$ orientifolds}

As we anticipated in Section 2, toroidal $\Omega$ orientifolds
are related by T-duality to toroidal $\Omega{\cal R}$ ones,
where K\"ahler and complex structures are interchanged.
However, for orientifolds of toroidal orbifolds the situation 
is a bit more involved,
as under T-duality, $(z_L,z_R)\to (\bar z_L,z_R)$, in general
the left-right symmetric $\bb{Z}_N$ action is mapped to a left-right
asymmetric $\hat{\bb{Z}}_N$
\eqn\asym{ 
(z_L,z_R)\to (e^{2\pi i\over N} z_L,e^{-{2\pi i\over N}} z_R)\ .
} 
Thus, in general T-duality relates symmetric with asymmetric orientifolds,
but for the $\bb{Z}_2$ orbifold the action \asym\ remains 
left-right symmetric. 
In this particular case T-duality identifies
\eqn\ztwo{  
{ {\rm type\ IIB}\ {\rm on}\ T^2\times T^2
             \over \{\Omega,\bb{Z}_2\}} 
}
with 
\eqn\ztwob{  
{{\rm type\ IIB}\ {\rm on}\ T^2\times T^2
             \over \{\Omega{\cal R},\bb{Z}_2\}} \ .
}
As a result, the two independent values of the quantized $B$-field for
the $T^2$ factors
in \ztwo\ are mapped to the corresponding ${\bf A}$ and ${\bf B}$ 
tori in Figs. 3 and 4. 
For instance, the $\Omega$ orientifold with vanishing $B$-field corresponds
to the $\Omega{\cal R}$ orientifold on the ${\bf AA}$ torus, as indeed
can be confirmed by a direct computation. 

Turning on a $B$-field, or equivalently rotating the cells, 
leads to the expected
rank reduction of the gauge group and introduces multiplicities
in the Neuman-Dirichlet sector, that are crucial for the cancellation of the
anomalies by a generalized Green-Schwarz mechanism \ggs. 
In \comm, these multiplicities had been related to the
(modified) structure of the fixed points in the presence of a
non-vanishing $B$-flux, and indeed the fixed points of the $\bb{Z}_2$
orbifold fill multiplets of dimension $2^{r/2}$. Thus, a
D5-brane in the presence of a rank $r$ $B_{ab}$
corresponds to $2^{r/2}$ (equivalent) copies, each sitting at a fixed
point in the multiplet, while the multiplicity $2^{r/2}$ in the 95
sector reflects the number of equivalent ways of building an $ND$ string.

An alternative geometric description can now be given in the context
of $\Omega {\cal R}$ orientifolds. Under T-duality, the D9 and D5
branes are mapped to D7-branes stretching along the horizontal and
vertical directions in Figs. 3 and 4, respectively. In the {\bf A}
torus the intersection number of these two D7-branes is one while
it is two for the {\bf B} torus\footnote{$^1$}{The relation between these
intersection numbers and the extra open string multiplicities $\kappa_k$
introduced in \bgk\ is explained in more detail in \bgkfut.}. 
Thus, in the $\Omega{\cal R}$ orientifolds,
the extra multiplicities in the 95 sector translate into 
the intersection numbers of the corresponding D7-branes: one for the
{\bf AA} lattice (corresponding to $r=0$), two for the {\bf AB}
lattice (corresponding to $r=2$) and finally four for the {\bf BB}
lattice (corresponding to $r=4$).

\vskip 36pt 

\noindent
{\bf Acknowledgments} We are grateful to L. G\"orlich, B. 
K\"ors and A. Sagnotti for interesting discussions.
C.A. would like to thank the Physics Department of 
Humboldt University for the warm hospitality. 
This work was supported in part by EEC TMR contracts ERBFMRX-CT96-0090
and ERBFMRX-CT96-0045.


\listrefs

\end